# Epileptic Seizure Risk Assessment by Multi-Channel Imaging of the EEG

Tiago Leal, Fábio Lopes, César Teixeira, António Dourado (*Senior Member, IEEE*)

CISUC-Centre for Informatics and Systems of the University of Coimbra
Department of Informatics Engineering
University of Coimbra, 3030-290 Coimbra Portugal

leal@student.dei.uc.pt   fadcl@dei.uc.pt   cteixei@dei.uc.pt   dourado@dei.uc.pt

*Abstract*— **Refractory epileptic patients can suffer a seizure at any moment. Seizure prediction would substantially improve their lives. In this work, based on scalp EEG and its transformation into images, the likelihood of an epileptic seizure occurring at any moment is computed using an average of the softmax layer output (the likelihood) of a CNN, instead of the output of the classification layer. Results show that by analyzing the likelihood and thresholding it, prediction has higher sensitivity or a lower FPR/h. The best threshold for the likelihood was higher than 50% for 5 patients, and was lower for the remaining 36. However, more testing is needed, especially in new seizures, to better assess the real performance of this method. This work is a proof of concept with a positive outlook.**

## I. Introduction

Epilepsy affects approximately 1% of the world's population. About 30% of these suffer from refractory epilepsy, for which neither surgery nor antiepileptic drugs are effective. These patients need another solution to improve their quality of life. The most researched solution is the prediction/forecasting in real-time of a seizure, raising an alarm or providing an occurrence likelihood by a transportable device that allows the patient to act in order to minimize negative consequences.

Electroencephalogram (EEG) is the most used signal in studies about epilepsy. To create a seizure prediction/forecasting algorithm, different brain states need to be labeled. The main ones (for prediction) are the pre-ictal stage – the period before the seizure – and the inter-ictal stage – the period where there is no seizure nor in the short future neither in the short past. The ictal stage is the one during which the seizure is occurring.

EEG signals are complex and sensitive to noise, artifacts, and other interferences. Seizure generation processes are different from patient to patient, which is an additional challenge to develop such algorithms, that therefore must be patient specific. Until today, after about 30 years of research, there does not exist yet an algorithm whose performance is reproducible in real life.

Previous seizure prediction research focused mainly on extracting features from the EEG and training a patient-specialized method, based on machine learning, thresholding techniques, etc.; see the reviews [1] [2] for a general overview. Then, to test it, several performance metrics can be computed, such as Sensitivity and False Positive Rate per hour (FPR/h).

In this work deep learning is used using the EEG signal without any feature previously extracted. There are a few studies using this approach [3][4][5], some reporting good results with the performance metrics they use. But the lack of access to a vast and continuous long-term EEG data, combined with the fact that deep learning methods need a great amount of data to be trained, caused the results not to be reproducible in real life. Using multi-channel EEG, transforming the multidimensional time-series into images, and using deep learning has recently been used for seizures detection [13].

The EEG time-series are directly transformed into images, which are used to train several Convolutional Neural Networks (CNN) to predict seizures. These CNN predictors will then be used to measure the likelihood that one seizure might happen at any time, i.e., acting as a forecaster, by considering the softmax layer output, that returns the likelihood of the image given to the CNN being of the inter-ictal or the pre-ictal class.

Likelihood thresholds were created over the mean of softmax layer outputs for the last minute of data (60 images) to measure the likelihood of an incoming seizure. The best threshold for each patient was then found by trial-and-error, in a testing set (not used in training the CNNs), by checking its ability to predict seizures.

## II. Methodology

EEG data of forty-one patients from the European Epilepsy Database [6] were used. All of them were pre-processed by the method explained in [7] and [14]. The EEG signals were filtered using a 0.5-100 Hz bandpass 4th-order Butterworth filter and a 50 Hz 2nd-order notch filter, to remove DC component, noise, and power line interference. The divided EEG segments were average referenced, and then analysed with extended-infomax Independent Component Analysis (ICA) from EEGLAB [12] was performed. The resulting independent components were visually inspected by experts to eliminate noisy ones. After that inspection, the signal was reconstructed using the non-noisy independent components. The EEG for each patient was composed by signals from 19 electrodes.

All patient's seizures were divided into two groups. The training group contains around two-thirds of the seizures and was used to train a CNN as a classifier. The second group was used for testing, but not in the usual way. Instead of that, it was used to find the best likelihood threshold by trial and error.

### A. Training

The reconstructed signal was transformed into images. First, normalization was applied. The whole signal (19 time series of the two groups of data) was centered in the mean and then divided by the double of the absolute maximum value of the signal, reducing it to the range [-0.5, 0.5]. Then, 0.5 was added to each of the values to turn them into the range [0,1]. The signal was then ready to be transformed into images.

Three types of images were created from the signal. The first type are the 1-second images. As the signal is composed by 19 electrodes sampled at 256 Hz, each 1-second image has dimensions 19x256. To create images for training, first the pre-ictal parts of the signal were divided into 19x256 segments by a sliding window, with 50% overlap, increasing the number of pre-icatal images to ease the class-balance operation. Then, the inter-ictal parts of the signal are also divided into segments of the same size by a sliding window, with the step of 1 second without overlap, since there are much more seconds of inter-ictal than of pre-ictal. From these, the same number of segments as the ones created in the pre-ictal parts were selected, to overcome the class imbalance problem. Furthermore, if for example 3000 inter-ictal images were needed, and there were 3 seizures, 1000 inter-ictal segments were selected from the interval before each seizure. This was done to increase the generalization capability of the CNN so it would not remain biased towards a particular seizure. The testing images were created by dividing the whole testing signal into segments with a sliding window of 1 second without overlap.

The two other types of images consisted in placing several (5 or 10) segments on top of each other, to create images that could encode implicitly time , expecting to encode brain dynamics. For 5-second images, the signal was divided by a sliding window of 19x1280 (1280 = 5 x 256) and for 10-second images, with a sliding window of 19x2560. These would then be divided into 5 (or 10) 19x256 images and placed on top of each other (with the earliest on top, and the most recent on the bottom). The 5-second images have then a dimension of 95x256 and the 10-second images of 190x256. An example of each kind of image can be seen in figure 1. The numbers 5 and 10 where chosem empirically.

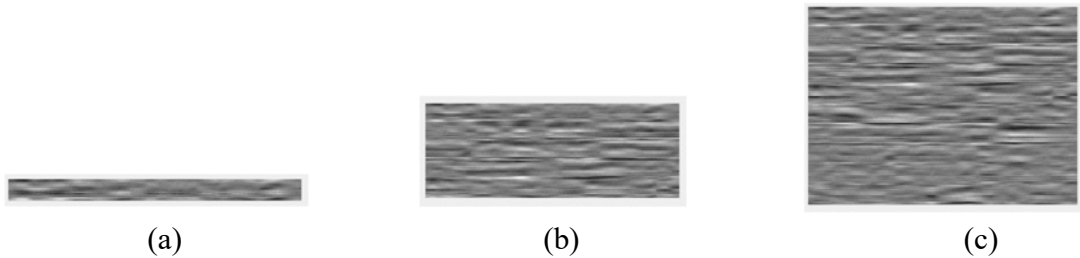

(a)                              (b)                              (c)

Figure 1- The three types of images created, 1 second (a) 19x256, 5 seconds (b), 95x256 and 10 seconds (c) 190x256

In the pre-ictal parts of the training signal, this sliding window moves with 50% overlap. So, the first image was composed by 5 (or 10) segments of 19x256 on top of each other, from the 0 seconds to 5 (or 10) seconds. The second image, from 0.5 seconds to 5.5 (or 10.5) seconds. The third, from 1s to 6s (or 11s) and so on. See the example of Fig. 2. For training inter-ictal images, similar segments were created from the training inter-ictal signal, but without overlap. The same number of pre-ictal and inter-ictal images was selected to overcome class imbalance, and for the inter-ictal the same number of images before each training seizure was randomly chosen. In the testing seizures, the whole testing signal is divided into segments with a sliding window of 19x1280 (or 19x2560) with a step of one second (no overlap) and transformed into 5-second (or 10-second) images.

Three types of CNNs were created, depending on the size of the images. All of them aimed to distinguish between inter-ictal and pre-ictal images, and were trained with 50 epochs, and stochastic gradient descent with momentum algorithm [11], implemented in Matlab Deep Learning Toolbox. A mini-batch size of 64 and a learning rate of 0.001 were chosen, the default values of the Matlab implementation [8]. Networks were trained for pre-ictal times of 10, 20, 30, and 40 minutes (these values where empirically chosen). So, for each patient, 12 networks were trained and tested, by combining the 3 types of images with the 4 pre-ictal times.

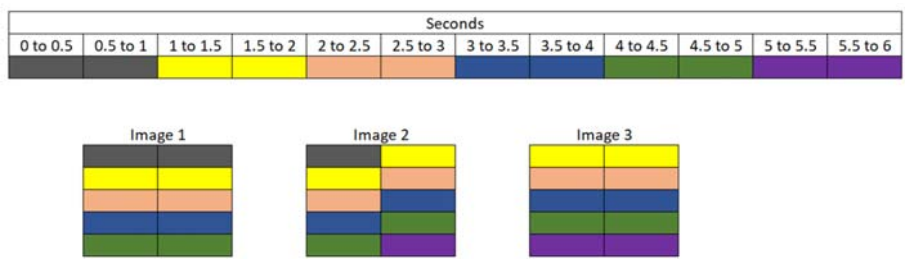

Figure 2 - The creation of 5-second pre-ictal images for training.

The CNNs used to classify each type of image are presented in figures 3, 4 and 5. Their parameterization has been made to take into account the different dimensions of the input images.

| The CNN for 1 second images has the layers: | The CNN for 5 second images has the layers: |
|---|---|
| Image input layer of 19x256. | Image input layer of 95x256. |
| Convolution layer with 64 3x27 filters, stride [1 3] and padding [0 1 1 1]. | Convolution layer with 64 3x11 filters, stride 2 and padding [1 1 2 3]. |
| Batch normalization layer. | Batch normalization layer. |
| Relu layer. | Relu layer. |
| Max Pooling Layer of 1x2 and stride [1 2]. | Max Pooling Layer of 1x2 and stride [1 2]. |
| Convolution layer with 128 5x5 filters, stride [1 2]. | Convolution layer128 5x5 filters with padding [0 0 0 1]. |
| Instance normalization layer. | Instance normalization layer. |
| Relu layer. | Relu layer. |
| Convolution layer with 256 5x3 filters, stride [1 2] and padding [0 0 1 2]. | Max Pooling Layer of 2x2 and stride 2. |
| Instance normalization layer. | Convolution layer with 256 5x5 filters, stride [2 3] and padding [0 1 1 1]. |
| Relu layer. | Instance normalization layer. |
| Convolution layer with 512 3x3 filters. | Relu layer. |
| Relu layer. | Convolution layer with 512 3x3 filters. |
| Dropout layer of 50%. | Relu layer. |
| Fully connected layer with 256 neurons. | Dropout layer of 50%. |
| Fully connected layer with 2 neurons. | Fully connected layer with 256 neurons. |
| Softmax Layer, which output is used to calculate risk. | Fully connected layer with 2 neurons. |
| Classification layer. | Softmax Layer, which output is used to calculate risk. |
|  | Classification layer. |

Figure 3. CNN for 1s images              Figure 4. CNN for 5s images

| The CNN for 10 second images has the layers: |
|---|
| Image input layer of 190x256. |
| Convolution layer with 32 3x27 filters, stride 3 and padding [4 4 1 1]. |
| Instance normalization layer. |
| Relu layer. |
| Max Pooling Layer of 2x2 and stride 2. |
| Convolution layer with 64 5x5 filters with stride 2 and padding [1 1 0 0]. |
| Instance normalization layer. |
| Relu layer. |
| Convolution layer with 64 3x5 filters. |
| Instance normalization layer. |
| Relu layer. |
| Convolution layer with 128 3x3 filters. |
| Relu layer. |
| Convolution layer with 256 3x3 filters. |
| Relu layer. |
| Dropout layer of 50%. |
| Fully connected layer with 256 neurons. |
| Fully connected layer with 2 neurons. |
| Softmax Layer, which output is used to calculate risk. |
| Classification layer. |

Figure 5. CNN for 5s images

*B. Finding the best likelihood thresholds*

For the seizures in the second group (the testing group), the whole signal was segmented into matrixes of 19x256, with no overlapping. After the segmentation, images of the three types (1, 5 and 10 seconds) were created from each and labelled according to the class they belong to. Note that the testing group was normalized as the training group.

Each image was given to the CNN of the respective size and pre-ictal time, and instead of the classification layer output, the softmax layer output was considered as a seizure likelihood occurrence indicator. This layer returns 2 values: the likelihood of the image belonging to the inter-ictal class, and the same for the pre-ictal class. Only the second value was considered because our aim is to measure the risk of coming a seizure.

Fig. 6*a* covers 60 s of one signal, i.e., the softmax output for 60 images of one second. The blue part represents the second output of the softmax layer for the 1s-images, and the black line the average of it for the last 60 images. Note that the black line continues what was happening before. The blue line connects each point with a straight line, by the plotting software. The black line is considered as the likelihood of a coming seizure.

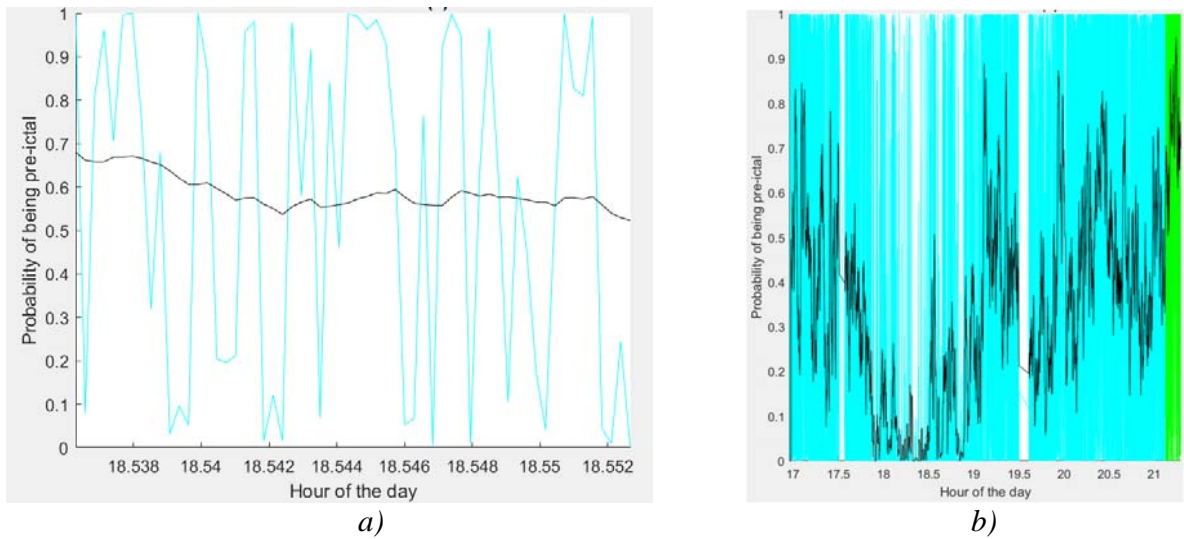

*a)*                                                                                                          *b)*

Figure 6 - Likelihoods given by softmax layer to each image being pre-ictal and the mean of the last minute (the black line); a) for a period of 60 seconds; b) four several hours, corresponding to the inter-ictal between two seizures s1 and s2 (in blue) plus the pre-ictal of the s2 (in green).

Figure 6*b* shows the softmax outputs for all the inter-ictal (in blue) and pre-ictal (in green) related to one seizure (the ictal time is not included, starts at the end of the image); this output varies considerably even for consecutive images (as seen on Fig. 6*a*). To overcome the difficulty that this causes, the mean of the last minute of probabilities was considered as in Fig.6a. The likelihood of an incoming seizure at one moment is measured by calculating the mean of the probabilities obtained for the last 60 images (the other 59 being the ones before the current). This value can be seen in Figure 6b by the black line. In the green zone of Fig. 6b) the likelihood is always over 0.5.

That likelihood was then used to predict incoming seizures. A threshold was created so that if the likelihood is over it, the current instance is considered to be of the pre-ictal class. Then, to predict, Firing Power technique was applied [10]. Only if more than $Y$ % of the images in last $X$ minutes were considered pre-ictal, the seizure alarm would be raised ($X$ being the pre-ictal minutes). $Y$ is the Firing Power threshold.

When raised, the Seizure Prediction Horizon (SPH) and the Seizure Occurrence Period (SOP) would follow [9]. The SPH is the period between the moment the alarm is given and the beginning of SOP. The SOP corresponds to the time interval in which it is expected that the seizure happens. The SPH was always 5 minutes, while the SOP was empirically fixed as half the considered pre-ictal time.

To summarize, images are given one by one to the trained network of the respective size and pre-ictal time. The softmax layer returns the probability of that image being pre-ictal. Together with the same probabilities of the last 59 images, the mean of the probabilities is computed. That mean is considered here the likelihood that a seizure could occur soon. To predict, a threshold $Y$ is needed such that, if in the last $X$ minutes, $Y$ % of the instances (means of the last minutes) are considered above it, an alarm is fired.

## III. Experiments

The objective of the experiments was to find, for each patient, which seizure occurrence likelihood threshold had the best performance in predicting seizures. This led to the test of several values for the best likelihood threshold $Z$ and the Firing Power threshold $Y$. So, for the likelihood threshold $Z$, empirical values from 0.05 to 0.9, with a step of 0.05 were tested. For $Y$, empirical values from 0.2 to 0.9 with step 0.1 were considered. Each of these were tested for different pre-ictal times and for each of the image types used.

For each patient, all combinations of image size, pre-ictal time, likelihood and firing power thresholds were tested, to find the best results, in a trial-and-error approach.

Table I shows the combinations of these variables that performed the best in predicting for each patient. Each seizure in which this test was done, had around 4.5 hours of inter-ictal and pre-ictal data. So, in Table I, 4.5 hours of testing group means that the test was done in one seizure, nine hours means it was done in two, and 13.5 hours in three.

## IV. Discussion

In 11 out of the 41 patients, the best likelihood thresholds can predict 100% of the seizures (Sensitivity = 1) with an FPR/h below 0.15, the maximum recommended by Winterhalder et al. [9]. Furthermore, in 26 out of 41 patients, that Sensitivity is obtained with an FPR/h below 0.3.

All optimal parameters vary considerably from patient to patient, which is something already known to happen in seizure prediction. The best pre-ictal time is 40 minutes in most patients. This could mean that, if tested with higher pre-ictal times, some results can eventually improve.

As for the image type, 1-second images produce the best thresholds in most patients. Such could happen due to the data overlapping caused by the technique to build the 5-second (or 10-second) images resulting eventually in CNN with lower generalization capabilities. In fact, this overlapping makes that the same half a second segments of the training pre-ictal signal appear in 10 or 20 images used to train, leading eventually to overfitting, which harmed the 5-second and 10-second images train and test. Nevertheless, some of those kinds of images still produced very good thresholds.

The values chosen for the thresholds for 36 of the 41 patients are lower than the 50% used by a normal classification layer. For the others it is over 50%. This means that using this method can help to predict some seizures that would otherwise go unpredicted, by lowering or increasing the threshold according to the patient. It can also be interpreted as some patients being more propense to suffer seizures when the likelihood is lower, while others can handle a higher seizure occurrence likelihood.

Nevertheless, for a considerable number of patients, the chosen thresholds are too low. Even if they performed well, when applied to new seizures there is a high probability that a high FPR/h can occur. Again, the importance of using more seizures to find the thresholds is highlighted. Using different types of seizures (some that happen when the patient is awake, others where he is asleep, for example) might help to improve these thresholds. Another solution is to keep updating the ideal threshold in real-time. That is, if a seizure occurs, use its data to adapt the ideal threshold to also consider its new information.

## V. CONCLUSIONS AND PERSPECTIVES

Using likelihood thresholds to predict helps in cases where the default 50% threshold used by a CNN classification layer may not be the most appropriate. By adjusting the likelihood threshold from patient to patient, some seizures that would otherwise be ignored may be predicted, and sometimes even the FPR/h can be reduced, as some patients may be more sensible to lower probabilities of having a seizure.

*Table I – Combination of values that had the best seizure prediction performance. 9h (13,5h) of testing time correspond to 2(3) seizures.*

| Patient Number | Pre-Ictal Minutes | Image Seconds | Likelihood Threshold Z | Firing Power Threshold Y | Sensitivity | FPR/h | Hours of Testing Group |
|---|---|---|---|---|---|---|---|
| 402 | 10 | 10 | 0,75 | 0,9 | 1 | 0,175 | 9 |
| 8902 | 40 | 5 | 0,15 | 0,85 | 1 | 0,064 | 9 |
| 11002 | 30 | 5 | 0,05 | 0,35 | 1 | 0,264 | 9 |
| 16202 | 10 | 1 | 0,6 | 0,2 | 1 | 0,028 | 13,5 |
| 21902 | 30 | 1 | 0,55 | 0,9 | 1 | 0,034 | 9 |
| 23902 | 10 | 10 | 0,15 | 0,9 | 1 | 0,134 | 9 |
| 26102 | 40 | 5 | 0,35 | 0,8 | 0,5 | 0,012 | 9 |
| 30802 | 40 | 1 | 0,3 | 0,7 | 1 | 0,255 | 13,5 |
| 32702 | 10 | 5 | 0,7 | 0,4 | 1 | 0,07 | 9 |
| 45402 | 40 | 1 | 0,1 | 0,75 | 1 | 0,3 | 9 |
| 46702 | 40 | 1 | 0,45 | 0,2 | 1 | 0,392 | 9 |
| 50802 | 10 | 1 | 0,25 | 0,45 | 1 | 0,294 | 9 |
| 52302 | 20 | 10 | 0,3 | 0,2 | 1 | 0,223 | 9 |
| 53402 | 20 | 5 | 0,3 | 0,6 | 0,5 | 0,054 | 9 |
| 55202 | 40 | 1 | 0,15 | 0,45 | 1 | 0,321 | 13,5 |
| 56402 | 20 | 10 | 0,05 | 0,3 | 1 | 0,24 | 9 |
| 58602 | 40 | 1 | 0,1 | 0,25 | 1 | 0,268 | 9 |
| 59102 | 20 | 5 | 0,3 | 0,45 | 0,5 | 0,013 | 9 |
| 60002 | 10 | 1 | 0,45 | 0,9 | 1 | 0,111 | 9 |
| 64702 | 40 | 1 | 0,15 | 0,25 | 1 | 0,28 | 9 |
| 75202 | 20 | 1 | 0,15 | 0,85 | 1 | 0,31 | 13,5 |
| 80702 | 20 | 5 | 0,05 | 0,4 | 1 | 0,121 | 9 |
| 81102 | 40 | 1 | 0,45 | 0,8 | 1 | 0,054 | 4,5 |
| 85202 | 30 | 5 | 0,3 | 0,25 | 1 | 0,245 | 9 |
| 93402 | 10 | 10 | 0,45 | 0,45 | 1 | 0,024 | 9 |
| 93902 | 40 | 1 | 0,3 | 0,7 | 1 | 0,216 | 13,5 |
| 94402 | 40 | 1 | 0,4 | 0,4 | 0,667 | 0,31 | 13,5 |
| 95202 | 10 | 1 | 0,3 | 0,2 | 1 | 0,192 | 13,5 |
| 96002 | 20 | 10 | 0,1 | 0,35 | 1 | 0,226 | 13,5 |
| 98102 | 40 | 1 | 0,3 | 0,65 | 1 | 0,248 | 9 |
| 98202 | 40 | 1 | 0,05 | 0,2 | 1 | 0,401 | 13,5 |
| 101702 | 40 | 1 | 0,05 | 0,65 | 1 | 0,301 | 9 |
| 102202 | 40 | 1 | 0,05 | 0,2 | 1 | 0,417 | 13,5 |
| 104602 | 20 | 1 | 0,3 | 0,5 | 1 | 0,226 | 9 |
| 109502 | 30 | 1 | 0,6 | 0,25 | 1 | 0,201 | 9 |

| 110602 | 20 | 1 | 0,1 | 0,5 | 1 | 0,349 | 9 |
|---|---|---|---|---|---|---|---|
| 112802 | 40 | 10 | 0,1 | 0,25 | 1 | 0,302 | 13,5 |
| 113902 | 40 | 1 | 0,05 | 0,2 | 1 | 0,356 | 13,5 |
| 114702 | 10 | 1 | 0,25 | 0,85 | 0,333 | 0,118 | 13,5 |
| 114902 | 40 | 1 | 0,35 | 0,2 | 1 | 0,296 | 13,5 |
| 123902 | 10 | 10 | 0,25 | 0,25 | 1 | 0,072 | 9 |

To find an ideal threshold that can be perfectly adjusted for a patient, a big number of seizures would be needed. An idea to overcome this problem may be to do more runs of this experiment, where the seizures used to train and test would interchange its role.

Calculating the threshold from which the patient is more prone to a seizure also helps to give more interpretability to the prediction, as CNN is a black-box classifier.

However, more testing is needed for this method. Other algorithms in different parts of the predictor should be explored to improve the performance of the likelihood thresholds. Also, more methods to calculate the likelihood, or more values to the different variables tested in the trial-and-error will make these tests more accurate. More methods to transform the signal into images should also be tried. Special techniques of image processing should be developed for these images from brain time-series.

The influence of the order of the electrodes when constructing the 5 s and 10s images should be investigated. Possibly it is not indifferent for the results since this order determines the spatial coverage of the image.

Another important question is to find out if with one single electrode, segmenting its time series and transforming each segment into one image, could be enough to obtain an acceptable performance. We are working on this idea.

## Acknowledgments

Tiago Leal acknowledges the 8 months research grant from CISUC, through the FCT - Foundation for Science and Technology, I.P., within the scope of the project CISUC - UID/CEC/00326/2021.